\documentclass{ws-ijmpcs}

\usepackage{graphicx}

\graphicspath{{pics/}}

\newcommand{\gs}{\gamma^\ast}
\newcommand{\lp}{l^+ l^-}
\newcommand{\Lagr}{\mathcal{L}}

\begin{document}

\markboth{S. Raspopov}
{Impact of $\gamma V$-vertex corrections on the $V P \gamma$ transition form factors}

%
%

\title{IMPACT OF $\gamma V$-VERTEX CORRECTIONS \\ ON THE
				$\omega \pi^0 \gamma$ AND $\phi \pi^0 \gamma$ TRANSITION FORM FACTORS}

\author{SERGII RASPOPOV}

\address{V.N.Karazin Kharkiv National University \\
4 Svobody Sq., Kharkiv, 61022, Ukraine \\
razserg@mail.ru}

\maketitle


\begin{abstract}
The aim of this paper is to present an effective field theory 
description of the conversion transition of the vector meson $V$ into
the pseudoscalar $P$ and the lepton-pair~$\lp$. The normalized form factor 
for $\omega\to\pi^0\gs$ transition is presented and compared
to the available data and to the predictions of other models.
\end{abstract}





\section{Theoretical Description}

We use the chiral Lagrangian in the vector formulation for spin-1 fields 
of Refs.~\refcite{Ecker:1988te}-\refcite{Prades:1993ys}.
In the even-intrinsic-parity sector one has
\begin{equation}
	\Lagr_{\gamma V} =  - \frac{e f_V  F^{\mu\nu}}{2} \bigl(
	\hat{\rho}^0_{\mu\nu} + \frac{1}{3}\hat{\omega}_{\mu\nu} -
	\frac{\sqrt{2}}{3}\hat{\phi}_{\mu\nu} \bigr)\, ,
\label{eq:lagr_gV}
\end{equation}
where $F^{\mu\nu} = \partial^\mu B^\nu - \partial^\nu B^\mu$, and
$\hat{V}_{\mu\nu} = \partial_\mu \hat{V}_\nu - \partial_\nu \hat{V}_\mu$ in terms of the vector fields 
for the vector mesons $V \equiv \rho^0, \omega, \phi$.
The interactions in the odd-intrinsic-parity sector read
\begin{equation}
	\Lagr_{V\gamma P} = - \frac{4 \sqrt{2} e h_V }{3f_\pi}
	\epsilon^{\mu\nu\alpha\beta} \partial_\mu B_\nu 
	\bigl( \rho^0_\alpha + 3\omega_\alpha + 
	3\varepsilon_{\omega\phi}\phi_\alpha \bigr)\partial_\beta\pi^0 \, ,
\label{eq:lagr_VgP}
\end{equation}
\begin{equation}
	\Lagr_{VVP} = - \frac{4 \sigma_V}{f_\pi}
	\epsilon^{\mu\nu\alpha\beta} \pi^0 
	\partial_\mu \bigl( \omega_\nu + \varepsilon_{\omega\phi} \phi_\nu \bigr) 
	\partial_\alpha \rho^0_\beta \, .
\label{eq:lagr_VVP}
\end{equation}
The $\epsilon^{\mu \nu \alpha \beta}$ is the totally antisymmetric
Levi-Civita symbol, pion decay constant is $f_{\pi} = 92.4$~MeV.
The terms with $\eta$ and $\eta'$ mesons and the G-parity-violating $\phi\omega\pi^0$ vertex are neglected.
The model parameters $f_V, h_V, \sigma_V$ are related via the special short-distance constraint:\cite{Prades:1993ys,Knecht:2001xc}
	$\sqrt{2} h_V - \sigma_V f_V = 0 \, $ .


The radiative decays of the light vector resonance $V$ into the pseudoscalar meson $P$
and photon are widely used as electromagnetic probes of the flavor content of the mesons.\cite{Achasov:2000wy}$^-$\cite{Usai:2011zza} 
For example, these decays provide an access to the value of the coupling constant $h_V$ via the partial width
\begin{equation}
	\Gamma (\omega\to\pi^0\gamma) = \frac{4\alpha M_{\omega}^3 {h_V}^2}{3f_\pi^2}
	{\biggl( 1 - {\frac{m_{\pi}^2}{M_{\omega}^2}} \biggr)} ^3
\label{eq:partial_width}
\end{equation}
Using PDG data\cite{Beringer:1900zz} for the $\omega \to \pi^0 \gamma$ decay we obtain $h_V=0.03753$.					

As it is seen from Eqs.~(\ref{eq:lagr_VgP})~and~(\ref{eq:lagr_VVP}), the transition $\phi \to \pi^0 \gamma$ is related
to the small parameter $\varepsilon_{\omega\phi}$, responsible for the $u\bar{u}+d\bar{d}$
component in the physical $\phi$ meson.	Thus one can compare the $\omega\to\pi^0\gamma$ and $\phi \to \pi^0 \gamma$ widths 
and find the value for the $\omega\phi$ mixing parameter
$\varepsilon_{\omega\phi}=(5.79\pm 0.17)\times 10^{-2}$.


The transition form factors can be extracted from the decay line shape $\frac{d\,\Gamma(V \to P \gs)}{d\,Q^2}$
and the cross section $\frac{d\,\sigma(e^+e^- \to \omega \pi^0)}{d\,Q^2}$.
Experimentally, only the normalized form factors are known $F_{V\to P\gs} (Q^2=0) = 1$.
					
We include direct $\omega \pi^0 \gamma$ coupling and mediated $\omega \pi^0 \rho$ with
subsequent $\gamma\rho$ conversion, contributing to the
Dalitz decay $\omega \to \pi^0 \mu^+ \mu^-$.
According to the Lagrangian terms from Eqs.~(\ref{eq:lagr_gV})-(\ref{eq:lagr_VVP}), 
the form factor reads
\begin{equation}
	F_{\omega \pi^0 \gs} (Q^2) =
	1 - \frac{\sigma_V f_{\rho}(Q^2)}{\sqrt{2} h_V} Q^2 D_{\rho} (Q^2) \, . 
\label{eq:from_factor}
\end{equation}						
An additional energy dependence of the EM coupling $f_{\rho}(Q^2)$
arises due to higher-order corrections\cite{Klingl:1996by};
$ D_{\rho} (Q^2) = \bigl[ Q^2 - M_{\rho}^2 - \Pi_{\rho} (Q^2) \bigr] ^{-1}\,$
is $\rho$-meson propagator.

\begin{figure}[h]
	\begin{minipage}[h]{0.6\linewidth}
		\center{\includegraphics[width=\linewidth]{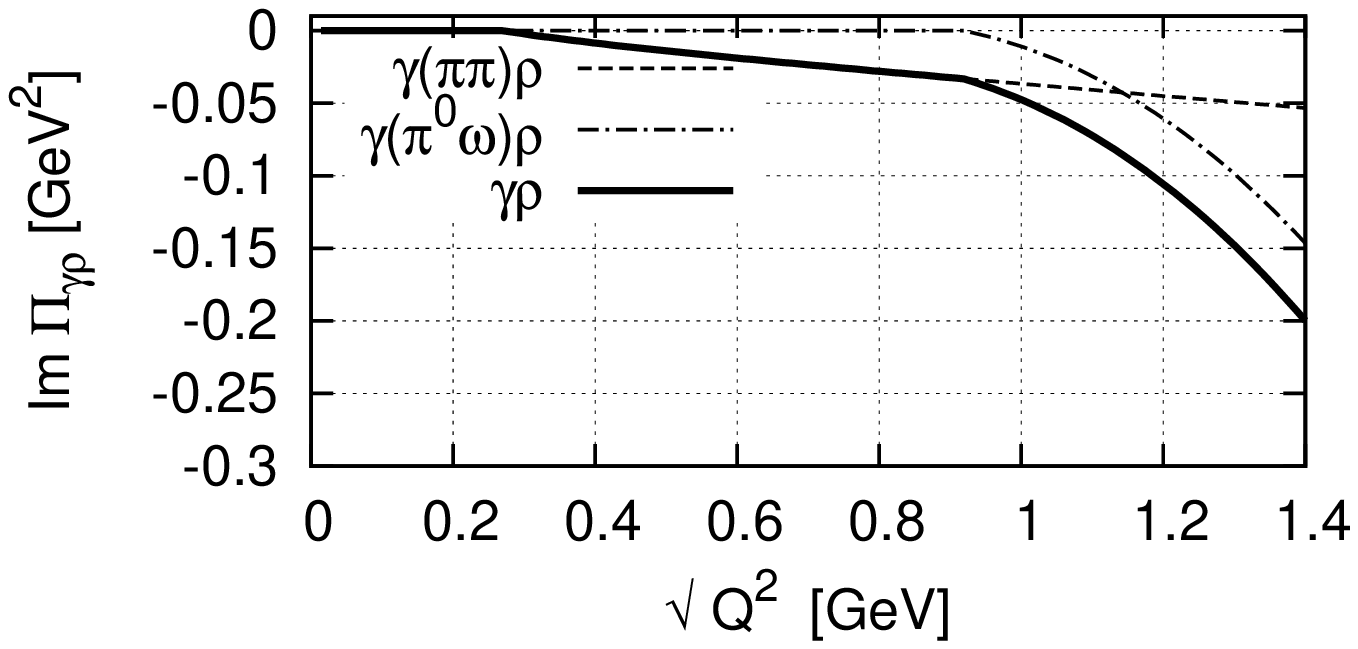}} 		
	\end{minipage}
	\quad
	\quad
	\begin{minipage}[h]{0.25\linewidth}
		\center{\includegraphics[width=\linewidth]{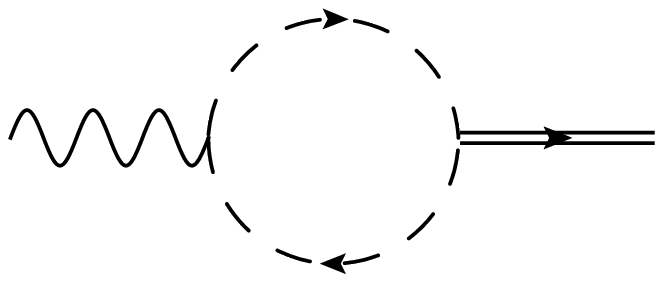}
		\includegraphics[width=\linewidth]{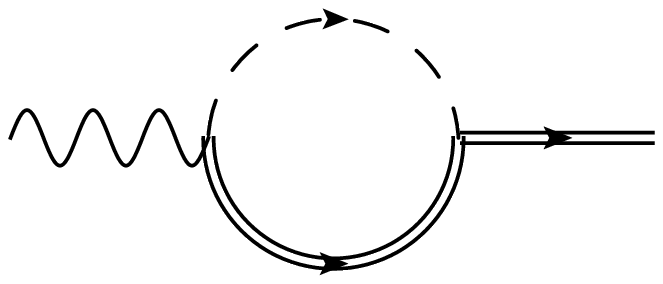}}
	\end{minipage}
\caption{The dominant loop contributions to the $\gamma\rho$ vertex.}
\label{fig:imP}
\end{figure}
					
In the region of interest the most important contribution 
to the self-energy operator $\Pi_{\rho} (Q^2)$ consist of the $\pi^+\pi^-$ and $\pi^0\omega$ corrections.\cite{Ivashyn:2006gf}
In the following we include only the imaginary parts of the loop contributions (see Fig.~\ref{fig:imP}).
This will be the dominant term for the energy-dependent width 
$\Gamma_{tot,\rho} (Q^2) = - M_{\rho}^{-1}\ {\rm Im}\, \Pi_{\rho} (Q^2) \, $,
which is important only for the $\rho$ meson within the scope
of current research in the region of momenta $0<\sqrt{Q^2}<1.4$~GeV overlapped with $\rho$ resonance. 
The ways to include the real part of the self-energy are discussed, e.g. in Ref.~\refcite{Klingl:1996by}.
					
The modified EM coupling in terms of the loop corrections
\begin{equation}
	f_{\rho}(Q^2) = f_{V} - \frac{\imath}{e\,Q^2}
	\sum_c {\rm\, Im} \Pi_{\gamma(c)\rho}(Q^2) \, ,
\label{eq:mod_coupl}
\end{equation}
where $c = (\pi\pi , \pi^0\omega)$ stands for the dominant loop contributions.
The ``bare'' constant $f_V$ is real-valued.
The modified coupling constant $f_{\rho}(Q^2)$ at $Q^2 = M_{\rho}^2$ has to describe
the leptonic decay width of $\rho$ meson: 
\begin{equation}
	\Gamma (\rho^0\to e^+e^-) = \frac{e^4 M_\rho} {12\pi} 
	{\bigl| f_{\rho}(Q^2 = M_{\rho}^2) \bigr|}^2 \ .
\label{eq:lept_dec}
\end{equation}
Eqs.~(\ref{eq:mod_coupl})~and~(\ref{eq:lept_dec}) allow us to find the bare coupling $f_V=0.202$.


\section{Results and Conclusions}

\begin{figure}[h]
	\begin{minipage}[h]{0.565\linewidth}
		\center{\includegraphics[width=\linewidth]{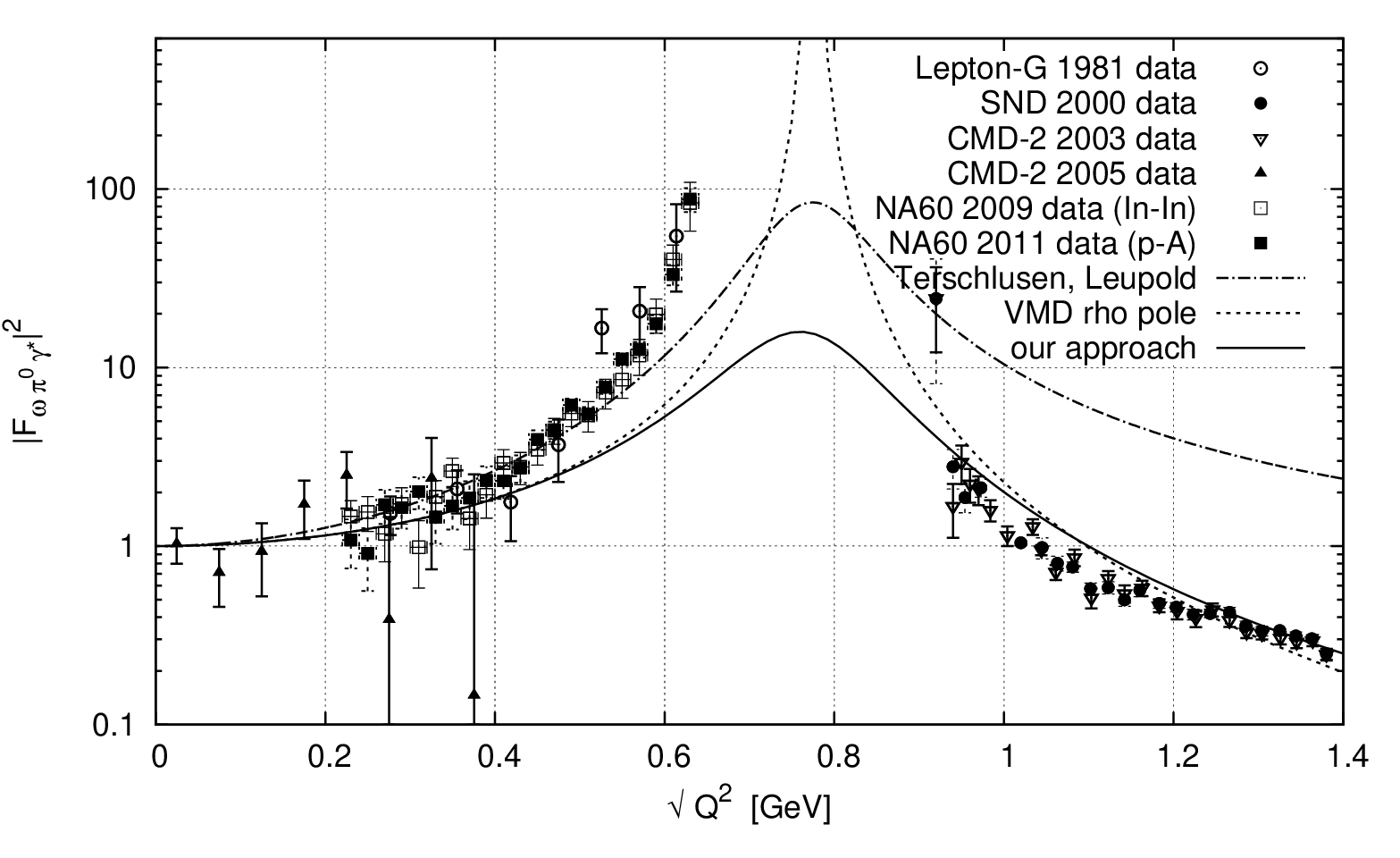}} 		
	\end{minipage}
	\quad
	\begin{minipage}[h]{0.39\linewidth}
		\center{\includegraphics[width=\linewidth]{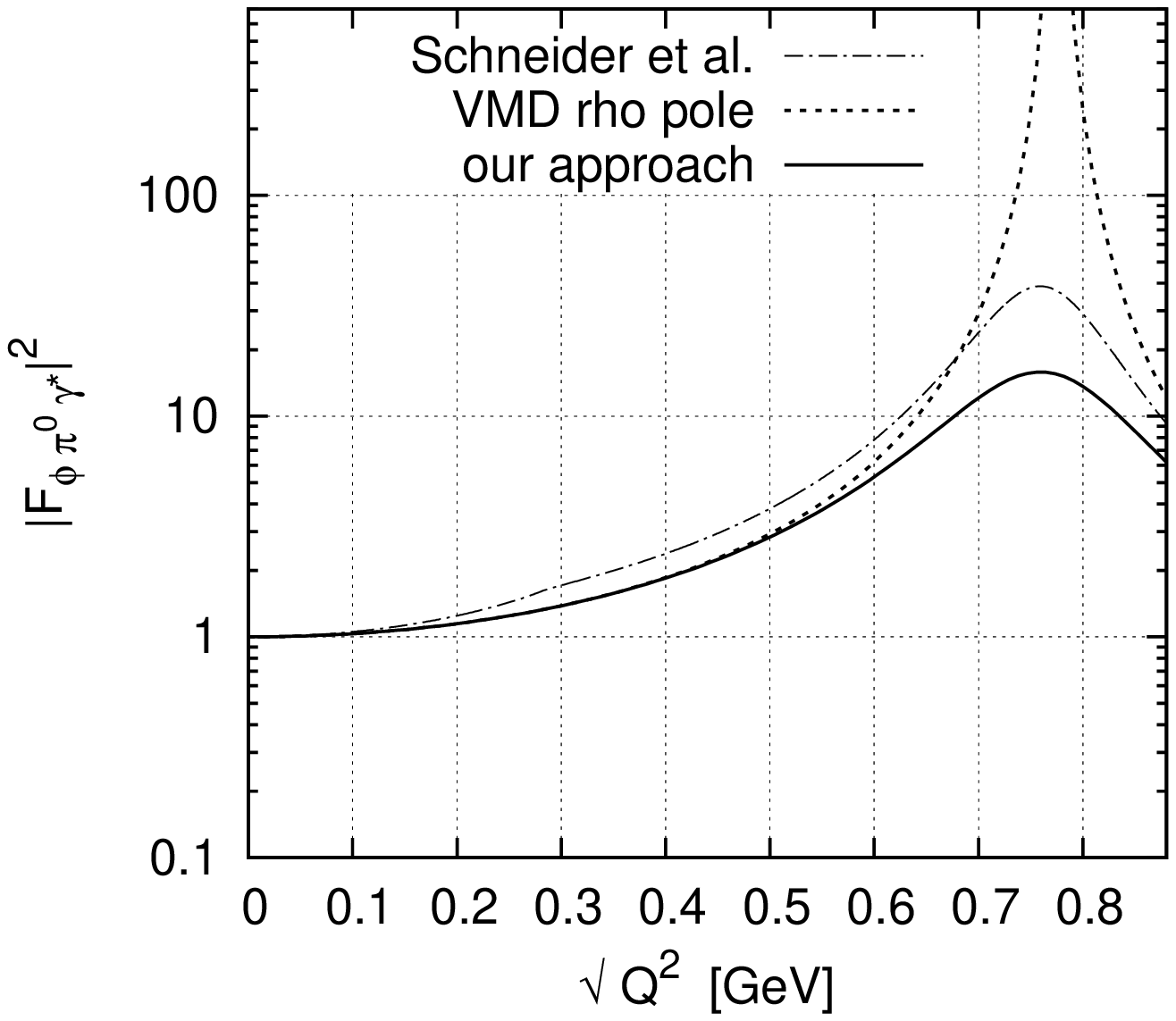}}
	\end{minipage}
\caption{The $\sqrt{Q^2}$ behavior of the $\omega \to \pi^0 \gs$(left) and $\phi \to \pi^0 \gs$(right) transition form factors.}
\label{fig:plot}
\end{figure}

The $\omega \to \pi^0\gs$ form factor (normalized) is shown in Fig.~\ref{fig:plot},~left.
We can notice that our model agrees with data slightly better than the model of Ref.~\refcite{Terschluesen:2010ik},
but for both models the problematic region is near $0.6$~GeV. 
This fact makes the problem of $F_{V\to P\gs}$ modeling very important. 
Also our results for the $\phi \to \pi^0\gs$ form factor (Fig.~\ref{fig:plot},~right) is compared with obtained 
in Ref.~\refcite{Schneider:2012ez}. This process is not measured yet and represents strong experimental interest 
(e.g., see materials by M. Mascolo in this proceedings).
Although the proposed approach is in qualitative agreement with the data from $e^+e^- \to \omega \pi^0$ at high energies, 
the $\gamma V$ vertex modification is not enough to reduce the discrepancy with NA60 data in the region $\sqrt{Q^2} > 0.4$~GeV.


\end{document}